# STUDY OF COMPUTER NETWORK ISSUES AND IMPROVISING DROP RATE OF TCP PACKETS USING NS2

Shweta Gambhir[1] and Kuldeep Tomar[2]

[1]Department of Computer Engineering, NGF Engineering College & Technology, Palwal, Haryana
[2] Department of Computer Engineering, MRIU, Faridabad, Haryana

## ABSTRACT

*As the enormous use of internet increases day by day so as security concern is also raise day by day over the internet. In this paper we discuss the network security and its related threats and also study the types of protocols and few issues related to protocols in computer networks. We also simulate the design of 5 node wired network scenario, its packet drop rate analysis through TCP protocol using NS2 as a simulator. Analyzed the performance of 5-node network when the packet is drop down by graphical method also called as Xgraph when rate parameter is in mb and also analyzed the performance of same network by changing the value of rate parameter at same time so no packets would drop down at same time and also analyzed the performance by Xgraph method.*

## 1. INTRODUCTION

In Today's world the use of Internet is growing considerably day by day in the scope and there are so many results revealed for the functional requirement of the internet which is in the form of number of new protocols and so many designed algorithms. These functional requirements include security, QOS (quality-of-service), policies, mobile networking etc. these algorithms and protocols growth gives the solution of so many problems for today's world of internet. Several protocols like SIP, HTTP, UDP, ICMP, TCP, RIP etc. expansion gives us privacy and also secure our data on the internet. In real world till now there are number of algorithms and protocols developed and designed (also by simulator) to cover the operational needs of internet but they certain have some shortcomings. There are so many simulator tools developed and evolved to test the work of real environment but they still have certain limitations because by using the different approaches such as the use of several topologies and another one is lot of traffic generation on the network make simulator task difficult. In about 1991 internet gains popularity and now become a part of our real world.

As we know that there are number of application users who used internet or they have number of so many task. Different internet user use the range of internet as for watching movie and Television, serials, news etc. by the help of many internet service provider(ISP) who offers the large range of IP addresses to their users by two mediums as telephone lines or by wireless network as USB. There are so many companies that provide these kinds of services and through these services user share their data and files over the internet, however for using these kinds of applications user needs range of high bits for transmitting large file, which consume network bandwidth. Although the ISP (internet service provider) has some limited range of IP address due to this reason some users share common or same bandwidth range on the Internet so this is the network where the demand is higher than the capacity. This causes network congestion or negative impact, on the data transmission rate and quality from the user perspective. The major part of the Internet is designed to forward data with equal priority, independent on the data source and destination.

## 2. LITERATURE REVIEW

As in [1] the study of TCP in NS-2 in which they worked on different formulae to calculate the performance of TCP in NS-2 and also gave installation steps for NS-2.we study about the 4 types of traffic and two protocols (TCP,UDP) how to create the topologies and network components. With NS-2 they design the wired network, wireless network and their performance analysis by graphical method [2]. As we know that the NS-2 is open source tool and widely used for network simulation area and there are different kind of traffic is generated for the different type of packets so this traffic and its parameters calculate the overall TCP/UDP packet drop rate by different methods such as by calculating the throughput by running AWK, perl script etc. in NS-2 as Dr. Neeraj Bhargava and Dr. Ritu Bhargava, Anchal kumawat, Bharat Kumar calculate the performance of TCP throughput by different parameters of simulation they use VBR traffic which is generate the video traffic at source to destination in form of packet as in [3]. They analyse the performance using both the tools Xgraph and Gnuplot. These both are the tools of NS-2 which shows the performance of network project graphically. We analysed the performance by Xgraph method in this paper. Up to this study of NS-2 and how well it works on the wired network and calculates its performance or throughput using different simulation tools and traffic generators. Now we will study about the integrating the agents and its combined security and integrity work on the wired, wireless and on sensor network using NS-2 as a network simulator they work well in the area of security how to encrypt and decrypt the data so it provide the security to data or information in the network [4].As so ever there are so many projects and research work is done under the area of wired and wireless network by using simulator NS-2.As NS-2 is also used as the REAL simulator in between year 1988-1990. NS-2 is widely used for security purpose and for doing simulation before using it as real world. We work in area of TCP packet when it will drop and when it will not drop in this paper using NS-2.

### 2.1. Comparative Analysis

Mr. Ajay Singh and Dr. Pankaj Dashore did well in their paper as in [5]. But they only analyze the packet flow, how packets come from source to destination through the router node and analyze its drop rate using Xgraph tool in NS-2 but we analyzed packet flow and drop rate as well as we improvise the packet drop rate we also explain how the packet drop rate is improvised and no packet would drop down at same time by altering the rate parameter of Exponential traffic.

## 3. TYPES OF NETWORK

### 3.1. Wired Network

Generally, physical connection between the two or more networks is a wired network. In wired network the data sent from one computer to another or one network to other networks through physical link or medium as ETHERNET cables etc. It is one way or two way interactions between two users and this is broadcast by wired method such as Ethernet cables or by optical cables etc. we can transfer the information from one computer to other via cables means data transmit by physical medium only. Security is key concerning issue over the internet, internet addicted world or in almost all organizations. There are so many methods introduced through which we will secure our network. What are the threats and how to overcome threats by given methods are discussed in section 4,5,7 and 8 how to secure the network through unintended user who access the authorized and authenticated information by different routing protocols mechanism or by other protocol blocking mechanism as from concern issue. We consider any of method which provides security to wired scenario and in Network Simulator NS2, with that methods information will be secured for wired and wireless network by generating different-2 types of traffic and their parameters.

## 3.2. Wireless Network

Internet is described as network of networks and Network defined as a network of devices through which two or more network interacts by using wireless technologies usually as mobiles, USB, Broadcasting medium. These are the some methods to provide interaction between two or more medium as Television, radio, cell phones. As cell phones called as wireless phones is a two-way interaction method for transmitting the data. Telecommunication is also the type to interact in the network. Wireless communication is the mean of transmitting the information and without the use of wires .Wireless Network communication refers to any type of computer or devices (for examples Access point, wireless Router) that is commonly associated for communications over wireless network to its interconnected nodes. This is also the Network security concerning issue all most all organizations which refers the wireless communication. Ad-Hoc network and Bluetooth are small type wireless devices through which data is transmit to one-to-many and it is also the security issue when someone hack the data via Bluetooth on the wireless network. The extensive fear over network is that when security is breaches due to the connectivity of one device to many on the network. Information security means that protecting information and information systems from any misuse, disruption, inspecting, eavesdropping, unauthorized access, perusal, recording or destruction disclosure, modification, Authenticating by unauthorized user. All these aspects are interrelating to each other and each of them has common goals and understanding of how to protect the integrity, confidentiality, availability etc. and however they have some differences in their work. Wireless Networks can be classified into some categories depending on different criteria (e.g. size of the physical area that they are capable of covering and domain of their use).The Wireless network work with almost many kind of network area as we know about them as DAN, WAN, PAN, MAN etc. The term computer security, network security, information/data security are used interchangeably in the computer world. The main focus is on the security of the data and provides privacy and confidentiality service to their end users. And NS2 is a best tool which provides security mechanism to wireless network and there are so many way through which network is secured on network.

## 4. SECURITY

All security system must provide a pack of security functions that offers the secrecy to the system on the network. The secrecy of data on the computer network is generally referred as security system. These goals can be listed in the following main categories: Integrity, Authentication, Confidentiality, Access control and Availability etc.

The main method is used to give security from Eavesdropping of data is Encryption which uses the cryptography algorithms [6] which is divided in two parts symmetric and second is asymmetric. There are so many methods or algorithms used today for encrypting or decrypting the data through which data/information is arranged systematically, and even through which data is encrypted one by one at sender side and same process of decryption of the data is also revised by same public key and different private key at receiver side. By using the different numbers of methods of cryptography we can secure our data. One is Symmetric which include DES [7], (RC5) [8] and AES [9] methods to encrypt data and to send that encrypted data securely on network and asymmetric techniques that are used to encrypt and decrypt the data by algorithm RSA[10], Diffie-Hellman [11]. Now a day's big data is also arranged on the internet and sharing of data is needed due to so many users or communication mechanism, privacy of data is also needed due to cyber attacks [12]. Cyber security, big data and its privacy are the major concern in the security region and there are so many researchers doing work on how to secure this big data and from cyber attacks.

## 5. ISSUES IN COMPUTER NETWORK

Recently in our world the environment of dynamic computing significantly increases very much in the scope. There are so many major attacks or issues of security are here in computer networks which would generally affect the security. As few of them would explain in this section:

**1.** Denial-of-service is a common type of attack which harms the security of network and decrease the performance of the router which involves the same IP address at source and destination port. So these types of attack lose the security and secrete information of user.
**2.** Another issue concerned with computer network is configuration if the configuration is poor or of low quality it will degrade the security.
**3.** Using many security policies at a time also affect the network and slower down the processing speed which also creates the problem in security.
**4.** Some malicious program can be the reason through which user can access information by creating a network access such as website.
There are few issues in network access through which information can be spoofed by unauthorized user.

## 6. STUDY OF PROTOCOLS

Different protocols used on the each layer of TCP/IP model which helps in communication from one network to more networks and also helpful to communicating with other layer protocols. As we know that first layer is physical and data link layer and many protocols are also used in this layer.

**1. Physical/ Data link layer:** In this layer data is transmit by wire or cables called as ETHERNET or as well as concerned with addressing the data from one network node to other, for this linking ETHERNET Cables are used. There are so many type of ETHERNET cables are For example: 10BASE5, co-axial, Twisted cable etc.
**2. Network layer:** Network layer tells how to route the packet on network for this purpose IP protocol used to route the packet from source node to destination node. The protocols are IP.IPX used at this layer.
**3. Transport layer:** To transport the routed packet that comes from the Network layer by means of protocols as TCP, UDP.
**4. Application layer:** Application layer used to display the data to the end user. There are so many protocols used at this layer for the end user through whom they can understand the data in user readable form so the protocols used at this layer are HTTP, SMTP, DNS etc.

## 7. ATTACKS ON NETWORK SECURITY

There are so many threats on network which will harm the security of user on the network we discuss few of them below in this:

**1. Intrusion Attack:** There are so many users on the network who use the internet and want security on their network but when any illegal person or unauthorized user attain access on the network to use the data of that user.

**2. Spoofing Attack:** When the illegal user access authorized information at transport layer by changing the source IP packet. These unauthorized users change the private information of the authorized user due to this the authorized user will receive altered information.

**3. Protocol based Attacks:** Communication protocols are the medium with the help of which data or information is transferred or retrieve on the computer network. There are so many attacks as discussed in next section come under this attack.

**4. Denial-of-service Attack:** DOS is a type of attack through which the network become weak and DDOS is distributed in the whole network and the appearance of the DDOS is present in the core architecture of internet [13] It is of two type[14]:

a. Ping of death
b. SYN attack

**5. Application level Attack:** Application level is called as user level where user get so many protocols which help them to retrieve the information from sender but attack is also generated at this level in form of malicious node Trojan on the HTTP etc.

**6. Logon Attack:** Due to security and privacy purpose logon policy is develop on the network but there are so many hackers who hack the password and username of the authorized user.

**7. Attack on address:** Address are of two types MAC address and physical address when some unauthorized user stole the information by accessing the address of source and destination.

There are so many researchers who researched the so many methods to detect the attacks or threats. As DDOS is a severe attack in network security there are few methods discussed in [15] [16] to detect the DDOS attack on the network.

## 8. PROTOCOL BASED SECURITY ISSUES

Protocol as we discussed above are the standard set of rules which is a mode of communication on the network. Each layer of TCP/IP model has some standard set of rules known as protocols, but sometimes what happen if some unauthorized or illegal user may be at anytime change the rules at any layer of TCP/IP model. Protocols mainly used to send their information in their communication mode to other next layer. As there are so many unauthorized or unauthenticated user who will alter the rules to steal the privacy and security of data. so as the matter there are few issues which will harm the security of protocols:

**1. Attack on ARP:** There are so many unauthenticated users who will attack on the physical address of destination so the source data will be send to wrong destination address.

**2. Attack on HTTP:** Trojan is malicious software on network which generated at application level on HTTP and IRC protocol to unsecure the data at user end [17].

**3. Attack on routing information protocol:** Routing the information means provide route to the data at IP layer. In this layer routing information is manually or dynamically handled by some protocols but some illegal user will change the information so the data is spoofed on the network [18].

**4. Attack on TCP:** Sometime any unauthorized user can predict the sequence number of the TCP protocol so that they can change the transferred information to other destination [18].

**5. Attack on SIP:** Session initiation protocol also used at application layer to provide the session limit facility to users on the network but when the messages or information is flooded or spammed so it can harm the security [19][20] so availability of data and its integrity may lost.

## 9. SIMULATION

Simulation is one of the important technology in this modern time. There are so many objects in real or hypothetical life. The network can also be simulated on the computer. A network simulator is a technique of implementing the network on the computer. Most of the simulators are graphical user interface driven, there are so many simulators which require commands or scripts as we talk about the NS2 which gives input in form of TCL scripts or in AWK scripts etc. The network parameters describe the state of the network (node placement, existing links) and the events. An important output of simulations is the trace file. Examples of network simulation software are ns2/ns3, OPNET, NetSim. Network simulators are particularly used to simulate, design and analyze the various kinds of networks, also analyze the performance by the effect of various network parameters.

### 9.1. Introduction to NS-2

For implementing the routing protocols, physical layer protocols (MAC, ETHERNET etc.), network protocols, transport layer protocols as being a simulator NS-2 play a fundamental role for work as a simulation tool for the different type network. NS-2 is a simulator that analyzes the performance of protocols and has number of protocols for implementing the network. NS-2 which describes the complete design of wired and wireless network, many security issues algorithms, for the queuing and routing queue mechanism of network. NS2 is very helpful because it is open source to design the new wired and wireless network for number of nodes and check the variability of that new designed networked algorithms and architectures and their network topologies. As we designed a 5 wired node scenario over UDP and TCP protocols and check their topologies, support many popular network protocols and also offer simulation results for wireless networks. As we know it support both wired and wireless network and suitable for the networking based projects and work on packets scenario how packet is transmit or receive from source to destination it also offer packet level inspection, how it is dropped and why it is dropped by using different traffic generators which generate traffic by using parameters. The Experimental traffic generator issued in this paper in which we setup a 5-node of wired network named tcp1, tcp2, tcp3 over source node 1, source node 2, source node 3 respectively in which the TCP agent connect to each source node as we know the we create a TCP agent before the packet/data sent from source to transmit the data. Because agent is needed through which the generated traffic will be transmitted the packet. So we generate the traffic at source with the help of agent TCP. There are different types of traffic generator which generate the traffic from source node they are CBR, Exponential, ftp, VBR etc. There are some steps which we discussed in section 10.1 for implementing the design of TCP protocol and analyze its performance when the packet will drop and when it will not drop.NS-2 Called as discrete event network simulator for simulating the IP. NAM is called as a Visualization Tool in NS-2 which gives the Window screen as an output [21] which shows the network scenario as in fig.3 called as NAM window which is resultant window screenshot for NS-2.

### 9.2. Architecture of NS-2

As we know NS-2 is a combination of C++ language and OTCL scripts i.e. object oriented tool command line interpreter which interpret the scripts line by line the code written in note pad or in ms-word etc. and save it by .tcl extension. There is a class hierarchy structure of NS-2 and extended class hierarchy which shows how to add components in NS-2 or to how to add new class because the NS-2 is a combination of TCL script and C++ language so we also write code in C++ language to run script in NS-2 [22].

International Journal in Foundations of Computer Science & Technology (IJFCST), Vol.4, No.4, July 2014

Design and run simulation by the Tcl interpreter using the simulator object in the OTcl library. The at most of the network components and the data path in event scheduler are implemented in the C++ and available to OTcl through an OTcl linkage library that is implemented using Tclcl, as there are two classes in Tclcl file one is class agent and second is class application.

**Class agents:** Each node in network that has to receive or sends the packets or in the terms called as traffic have an agent attach to the node so they will send or receive the data i.e. agent is needed, in this we create a TCP agents in our work.

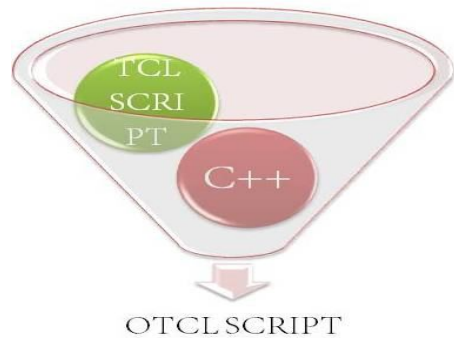

Figure 1: OTCL Script is a combination of tcl and C++

## 10. EXPERIMENTAL SETUP OF TCP

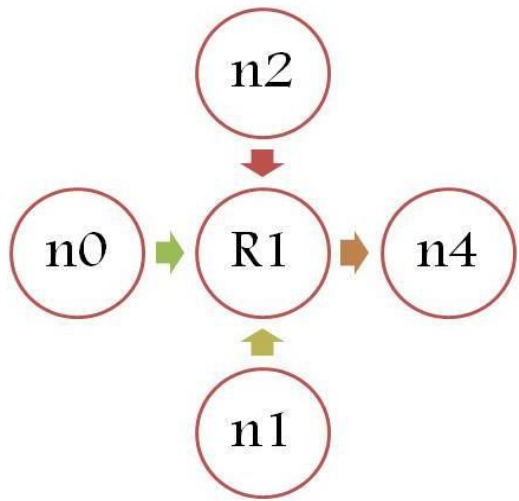

Figure 2: Experimental daigram of 5- node wired TCP network

The figure mentioned above is a experimental setup of 5 nodes in which n0,n1,n2 are the source node which generate packets by the agent TCP and exponential packet is attached to the three source nodes. The exponential traffic has 4 parameters :

1. Packet size
2. Rate
3. Burst time
4. Ideal time

These are the four parameters of the exponential traffic through which traffic is generated on source nodes and this traffic in form of packets send to router node r1 at that time due to congestion window or the packet rate the ssome data is fall down through the router node r1 and packets which are not fallen dowm are send to destination node n4.

## 10.1. How setup is done so that packets will send from source to destiantion(script)

Following are steps/script steps of how experimental setup of TCP is established and how packets are send from source to destination are given below:

1.Create the object of simulator first as set tc[new simulator].
2.Set the nodes and established a link between the nodes for this purpos we use routing queue to manitain the packet flow in queue as set node_name[$ ns node] and link may be simplex or duplex as $ns simplex_link $sourcenode $routernode mb ms RED/DropTail. And fb is a instance of simulator. mb is bandwidth is also the main cause to drop the packets & ms is delay between the packets.
3.Create a TCP agent at source because the TCP agent is created before packets sent from source as set tcp0 [new Agent/TCP] $tcp0 set class_ 2 $tc attach-agent $source node $tcp agent set sink0 [new Agent/TCPSink] $ns attach-agent $destination node $sink0 $ns connect $tcpagent $sink0.
4.Now attach the traffic given in this paper i.e. exponential traffic which is on/off distribution traffic.attach to the TCP agent, this traffic with their parameters attached. As set exp0 [new Application/Traffic/ Exponential].
$exp0 attach-agent $tcpagent $expotraffic set packet_size_ 210 $expotraffic set burst_time_ 2ms $expotraffic set idle_time_ 1ms $expotraffic set rate_ 100k.
5.Now attach the source traffic to sink, sink is a previous traffic which means sink has agent/loss monitor at the destination and monitor the amount of data byte sent or received so, amount of data lost which will calculate the bandwidth of nodes as $ set bandwidth number[$ sink set byte] code write on the script.
6. Out.tr.w is an output file which shows the output bandwidth of source node in form of graph called as Xgraph.
7. Run the script by saving extension .tcl and run the script. When script runs it shows the out.nam file which is called as network animator screen.
8. Now call the finish procedure in which close the o/p files as close $fo.

And then execute the Xgraph in geometry of window size it may be of 600 * 400. Now how to calculate the bandwidth for the network, for this firstly set time as set time=0.1, 0.2, 0.3 etc. Recall the procedure record at after the set time and also set now time to call the procedure again and again. Now record the time and also set the start time of each simulated source then set the end time of each source through which simulation can start and stop. Following mentioned above are some steps for how to write a script for analyzing the packet flow of network, how to recognized and also analyze its performance by graph which is shown in next section of this paper. In next section we analyzed some results for packet flow of TCP, in section 10.2 we analyzed that at any time the packet will drop through router when rate is in mb and analyzed its performance using Xgraph but we also discuss in section 10.3 results which analyzed that at same time instant not even a single packet is drop by router by altering the value of rate parameter in K unit.

## 10.2. Result Analysis When Packets Are Drop Down Through Router

In fig.3 to fig.6 shows that packets comes from three sources n0, n1, n2 that are transmitted to the destination n4 which is in hexagonal shape (in fig. 3) and packets are drop down at the router node called n3 which is in square shape (in fig. 3) . TCP agent is created at three sources and exponential traffic is attached to the agent but at that time rate parameter value is in mb so that packets would drop down at router node as shown in below figure.

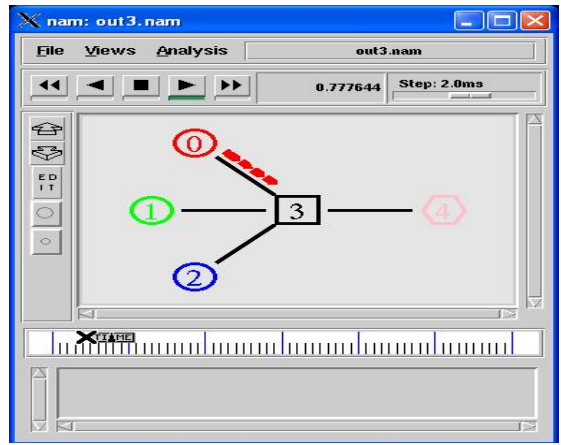

Figure 3: when packets send from source n0 to destination n4 through n3 (router node)

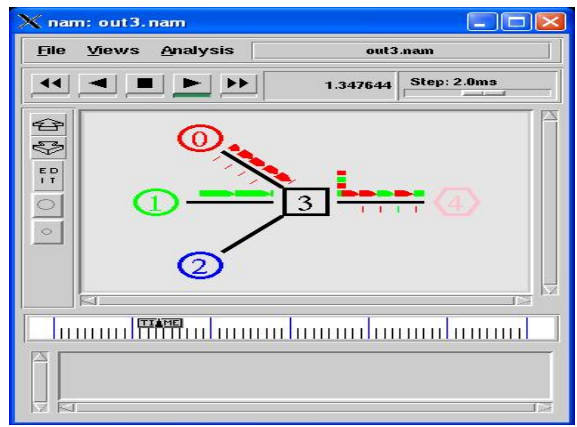

Figure 4: when packets send from source n1 to destination n4 through n3 (router node)

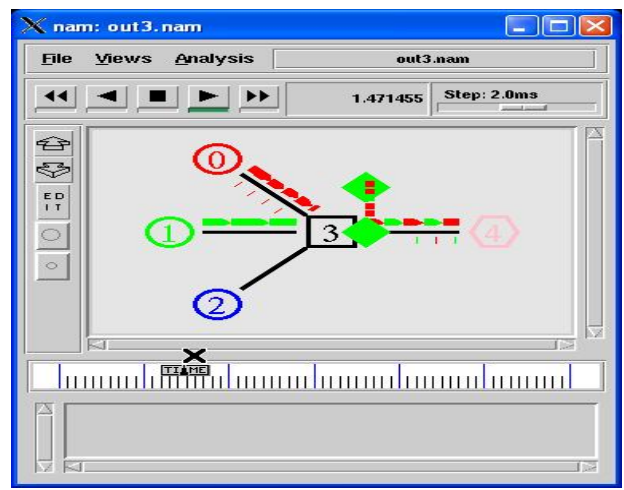

Figure 5: some packets of source n1 drop down at n3

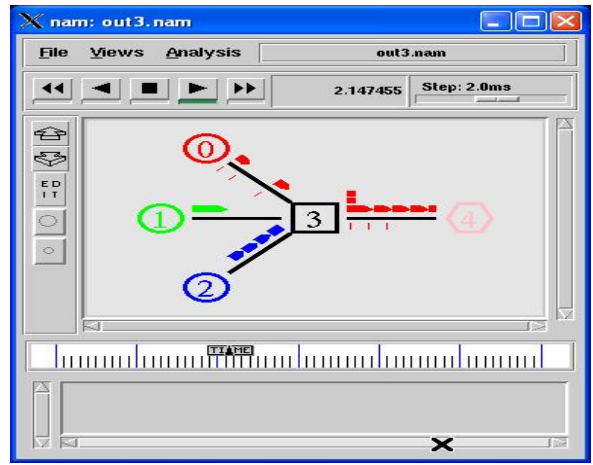

Figure 6: when packets send from source n2 to destination n4 through n3 (router node)

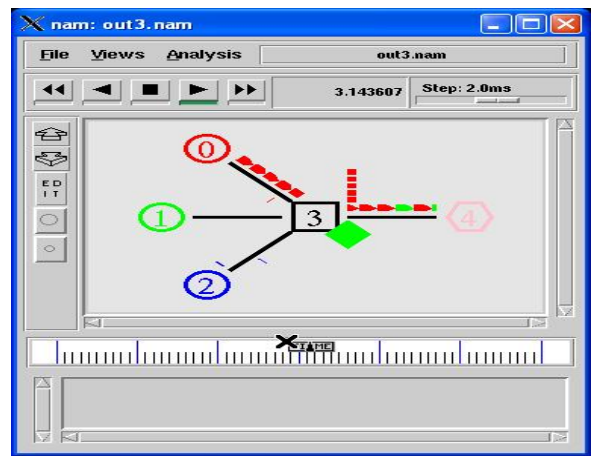

Figure 7: at time 3.14 packets would drop down when rate is in mb at n3

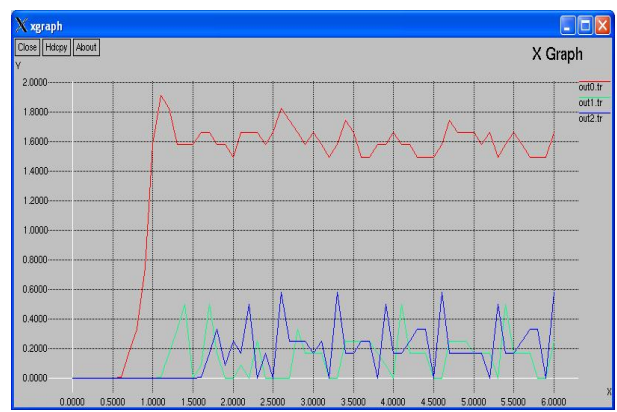

Figure 8: Xgraph shows the performance by X-axis which represents rate and Y-axis shows the time so the peak burst rate is in mb/sec. graph shows when packets drop at router node. At 3.14 time packet would drop down so at that time peak is at low.

## 10.3. Result analysis when packets are not drop down through router (improvisation)

In this section there is no packet would drop down at router node due to rate parameter is in K of exponential traffic as shown below:

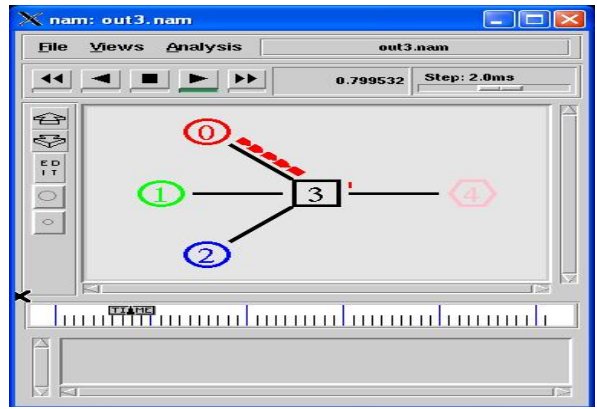

Figure 9: when packets send from source n0 to destination n4 through n3 (router node)

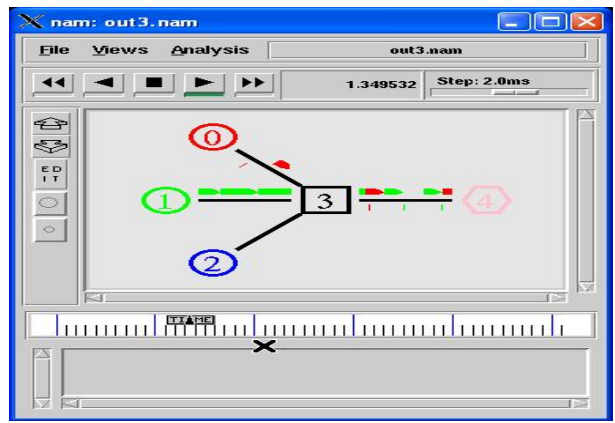

Figure 10: when packets send from source n1 to destination n4 through n3 (router node)

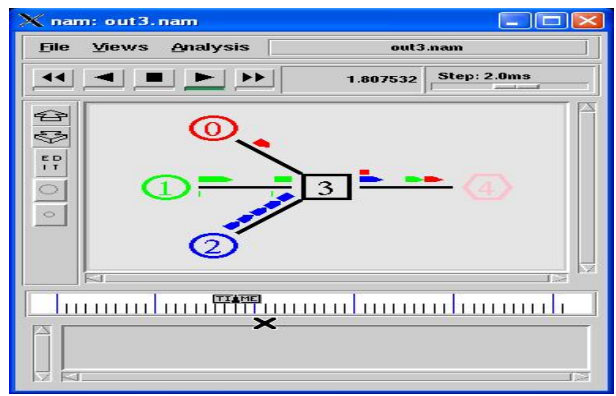

Figure 11: when packets send from source n2 to destination n4 through n3 (router node)

International Journal in Foundations of Computer Science & Technology (IJFCST), Vol.4, No.4, July 2014

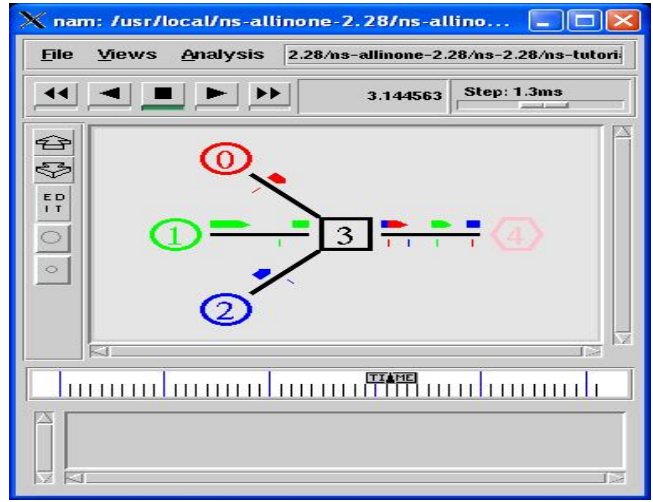

Figure 12: at time 3.14 packets would not drop down when rate is in K at n3

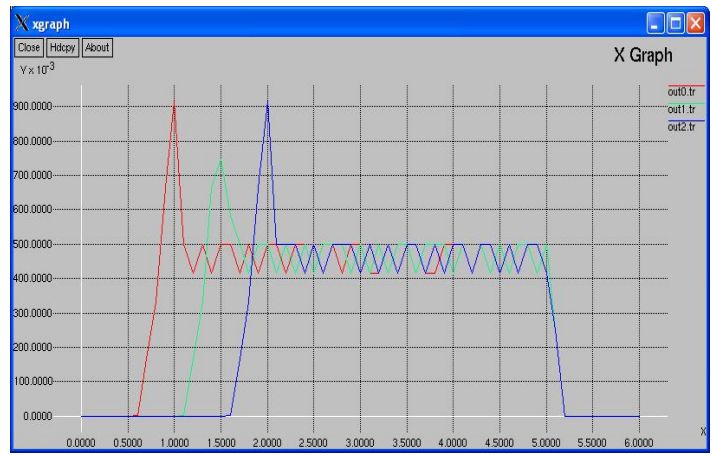

Figure 13: Xgraph shows the performance by X-axis which represents rate and Y-axis shows the time so the peak burst rate is in K/sec. graph shows when packets drop at router node. At 3.14 time packet would not drop down in this graphical result so at that time peak is at high at that time.

As the following results shown in the figure 7 the packet is drop at time instant 3.14 when parameter of exponential traffic rate is in mb and figure 12 shows the result of same time when the rate is in K so that time packet would not drop down and also analyzed graphical results which represents at 3.14 time packet would drop down so at that time peak is at low in figure 8 and result analyzed for in fig. 13 at 3.14 time packet would not drop down in this graphical result so at that time peak is at high at that time. So we analyzed the packet flow of TCP using the simulator NS-2. And above sections we study the computer networks its protocols and threats in network and in the protocols.

## 11. CONCLUSIONS

In this paper we discuss the network, its scenario and also its concerning issues. We also discuss the types of attack which affects the network so that results will the unintended user access the authorized data by different methods as discussed in section 7. We also study the types of protocols used on each layer of TCP/IP model and its related issues by means of which data is hacked or alter by unauthorized user by illegal way from sender side, at the router side or even at destination by attacking on the number of protocols used at each layer. We also analyzed the packet flow scenario i.e. how packet is flow from source to destination and how much packets are dropped at router node by rate parameter of exponential traffic. As exponential is on/off traffic at on time packets are dropped and at off time packets would not drop, each parameter has different effect so on changing the rate parameter of exponential traffic would affect on the packet drop rate and bandwidth is also affecting the packet drop rate of the network. We create only TCP agent because as we know that TCP is connection oriented protocol and connection is established before packets flow from source to destination and acknowledgement is also send by receiver. So as less packet drop rate count as comparison to UDP protocol. But some packets at TCP would loss because of the rate parameter is in discussed in section 10.2 but when we alter the value of rate parameter as in section 10.3 no packet would drop by router, all the packets are sent to the destination n4 form the source n0, n1, n2.

## Authors

Ms. Shweta Gambhir is a M.Tech Scholar in the department of CSE at NGF College of engineering & Technology, Palwal, Haryana, India. She did her B.Tech in CSE in 2011 from WIT College affiliated to MDU Rohtak. She also has published papers in national and International conferences/ journals.

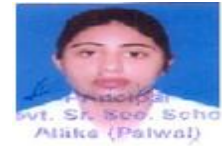

Kuldeep Tomar is a Research Scholar in the Department of CSE, MRIU, Faridabad, 121001, India. Ccurrently working as Head of the Department of CSE in NGFCET, Palwal, Faridabad, Haryana. He was born in Sonepat, Haryana on 2nd Oct, 1978. He has done M.E/M.Tech in Computer Science and Engineering from C.I.T.M., Faridabad, India.He has a total experience of 12 years in different organizations. He is currently working as Associate Professor in NGF College of Engineering & Technology, Palwal, Faridabad, Haryana. He also has worked as Assistant. Professor in Skyline Institute of Engineering & Technology, Gr. Noida; as Senior Lecturer in B.S.A.I.T.M., Faridabad; as Technical Head/Manager at SSI (Software Solutions Integrated Ltd. and as Sr. Faculty at Hartron Workstation (Haryana Govt. Undertaking).He has published more than 12 papers in International/National Journals and conferences etc. Has is also written a book. He also is a member of Computer Society of India, Membership No:N1039627.

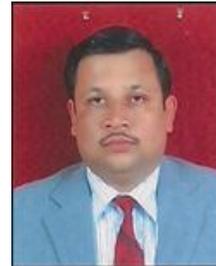